\documentclass[
groupedaddress,
twocolumn, 
notitlepage,
bibnotes,
showkeys,
aps, pra, 10pt]{revtex4-2}

\usepackage{graphicx}
\usepackage{physics}
\usepackage{amsmath}
\usepackage{amssymb}
\usepackage{cancel}
\usepackage{empheq}
\usepackage{xcolor}
\usepackage[colorlinks,citecolor=blue,urlcolor=blue,linkcolor=blue,hypertexnames=true]{hyperref}
\usepackage{tabularx}
\newcolumntype{A}{>{\centering\arraybackslash}X}
\usepackage{array, makecell}
\usepackage[english]{babel}

\newcommand{\eg}{\textit{e.g.}~}

\newmuskip\pFqmuskip
\newcommand*\pFq[6][8]{%
  \begingroup 
  \pFqmuskip=#1mu\relax
  \mathchardef\normalcomma=\mathcode`,
  \mathcode`\,=\string"8000
  \begingroup\lccode`\~=`\,
  \lowercase{\endgroup\let~}\pFqcomma
  {}_{#2}F_{#3}{\left[\genfrac..{0pt}{}{#4}{#5}~\middle|~#6\right]}%
  \endgroup
}
\newcommand{\pFqcomma}{{\normalcomma}\mskip\pFqmuskip}

\newcommand{\aell}{\abs{\ell}}

\begin{document}

\title{
\textbf{Near-focus intensity evolution of spatiotemporal optical vortex pulses \\ and consequences for their nonlinear upconversion
}
}
\author{Titouan Gadeyne}
\email{titouan.gadeyne@gmail.com}
\author{Vartika Vishnoi}
\author{Thierry Ruchon}
\email{thierry.ruchon@cea.fr}
\affiliation{%
Universit\'e Paris-Saclay, CEA, LIDYL, 91191 Gif-sur-Yvette, France
}

\begin{abstract}
Spatiotemporal optical vortex (STOV) laser pulses are attracting attention for their puzzling transverse orbital angular momentum (OAM) and propagation properties, which could determine their future applications. While the characteristic reshaping of their space-time envelope upon diffraction from the focal plane to the far-field is now well-established, the evolution of their intensity as they propagate has not been studied quantitatively, despite growing interest for their use in nonlinear optics and strong-field physics. Here, we explore experimentally and analytically the focusing dynamics of STOVs produced by a pulse shaper, and evidence a substantial longitudinal shift between the coordinate at which a circularly-symmetric STOV forms and that at which the pulse reaches its peak intensity. To illustrate some practical consequences for applications, we experimentally demonstrate how the yield of second-harmonic generation can consequently be maximized several Rayleigh lengths beyond the plane where the circular STOV really occurs. These results signal careful characterization of spatiotemporal pulses as a requisite for nonlinear optical experiments, and pave the way for finer control of complex laser drivers for light-matter interactions.
\end{abstract}



\maketitle

\section{Introduction}

Spatiotemporal optical vortex (STOV) pulses are laser wavepackets featuring a singularity in a space-time plane, around which their phase winds by an integer number $\ell$ of $2\pi$ cycles \cite{hancock_2019_free, chong_2020_generation}. Alike the more common vortex beams possessing a purely spatial phase vortex \cite{allen_1992_orbital}, STOV pulses also exhibit an orbital angular momentum (OAM) vector, but oriented perpendicular to their propagation direction. Owing to this unique transverse OAM, STOVs are expected to deepen our understanding of light-matter interactions: experiments have already investigated their perturbation by a transient plasma \cite{hancock_2024_spatiotemporal, hancock_2025_transverse}, and their nonlinear upconversion was recently achieved, via second-harmonic generation (SHG) \cite{gui_2021_second, hancock_2021_second} and via the strong-field process of high-harmonic generation \cite{fang_2021_controlling, martinhernandez_2025_extreme, gadeyne_2026_controlled}. Meanwhile, theoretical studies have kept flourishing, predicting intriguing behaviour in varied contexts including spin-orbit interaction under strong focusing \cite{bliokh_2021_spatiotemporal}, Wigner delays in reflection \cite{mazanov_2022_transverse}, plasma-mirror relativistic harmonic generation \cite{zhang_2022_intense, chen_2022_relativistic}, their role as dressing fields in photoionization \cite{chen_2023_atomic}, coupling to surface modes \cite{rodriguezfortuno_2025_bulk}, and more.

\begin{figure*}
        \centering
        \includegraphics[trim={0, 8pt, 0, 0}, width=\linewidth]{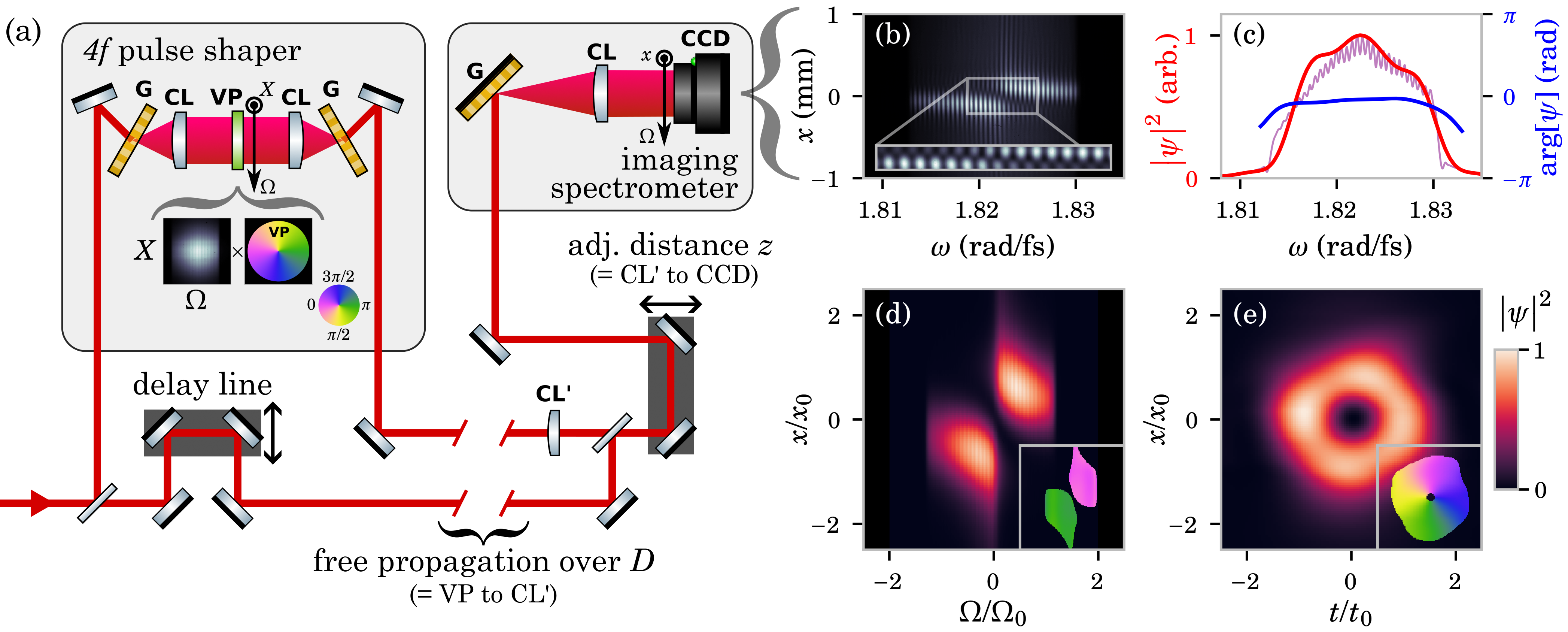}
    \caption{(a) Sketch of the optical setup (see text). The distance $D$ is counted from the vortex plate (VP) to the focusing cylindrical lens (CL'). The distance $z$ is counted from CL' to the CCD camera. (b-e) Example of a spatially-resolved FTSI measurement of a STOV pulse. (b) Raw spectral interferogram between the $\ell = 1$ STOV and the reference pulses. The characteristic $\pi$ phase shift between the two intensity lobes is highlighted in the inset. (c) spectral intensity (red) and phase (blue) of the reference pulse, retrieved from a FROG measurement. For comparison, the purple line shows the spectral intensity measured with our (higher-resolution) imaging spectrometer. (d) Spatiospectral complex pulse retrieved from (b) and (c). The phase is shown in inset and blanked where intensity falls below $10$~\%. (e) Spatiotemporal pulse envelope (intensity, phase in inset), obtained by a Fourier transform of (d).
    }
    \label{fig1}
\end{figure*}  

STOVs and transverse OAM however come with many theoretical subtleties, that have been vividly discussed in recent years (see \cite{hancock_2021_mode, bliokh_2023_orbital, porras_2024_clarification, gadeyne_2025_energy} and references therein), across studies that unearthed fundamental differences between the behaviour of spatial and spatiotemporal vortex pulses. A major aspect in which the two differ is in their stability upon propagation: while a spatial vortex beam retains its circular shape as it propagates, a STOV pulse does not and its space-time envelope instead breaks into several intensity lobes away from the focal plane \cite{hancock_2019_free,
hancock_2021_mode, 
porras_2023_propagation}. As the space-time field experienced by a target will depend on where exactly it is placed along the beam path, this characteristic reshaping of STOV pulses as they propagate is expected to make any interaction with matter strongly dependent to longitudinal positioning. Thus far, this reshaping has only been described from a \emph{modal} point of view, with an emphasis on the duality between space-time Laguerre-Gaussian (LG) and ``tilted'' Hermite-Gaussian (THG) modes, that convert into one another between near- and far-fields \cite{hancock_2021_mode, porras_2023_propagation,
porras_2025_optics}. Both experimental and theoretical studies have disregarded how this reshaping affects the intensity distribution within a STOV pulse from a quantitative standpoint, and what the consequences may be for applications, particularly in nonlinear and/or strong-field physics.

In this work, we investigate analytically and experimentally the reshaping of STOVs produced by a $4f$ pulse shaper near their focal plane, and how the peak intensity reached by these pulses evolves along the beam path. We discover a non-trivial dependence of this intensity, stemming from two factors. On the one hand, the known reshaping of STOVs from LG- to THG-type modes changes how concentrated the energy is within the pulse envelope: we observe the geometrical focus (the coordinate at which the circular STOV is formed) to actually be a local \emph{minimum} of the peak intensity. On the other hand, an experimental setup always comes with some finite propagation step between the shaping and focusing optics, playing no role in most cases. For a STOV pulse, we demonstrate how a substantial free-propagation distance pushes the embedding Gaussian caustic downstream of the geometrical focus (where the circular STOV still forms), shifting further away the coordinate at which the peak intensity is reached. We illustrate how this affects STOV-driven nonlinear processes, by measuring the yield of second-harmonic generation (SHG), which appears to peak several Rayleigh lengths downstream of the circular-STOV plane. These observations generalize to other spatiotemporally-structured pulses, and signal the need to carefully characterize the propagation of such laser drivers.

\section{Experimental setup}

The experiment was carried out on the ATTOLab platform at CEA Saclay. We used an Yb-based laser source delivering near Fourier-limited pulses around a wavelength $\lambda_0 = 1033.6$~nm ($k_0 = 2\pi/\lambda_0$, $\omega_0 = ck_0$). To shape and characterize STOV pulses, we built the setup sketched in \autoref{fig1} (a). The beam is first split into a Mach-Zehnder interferometer, and STOVs are shaped in the upper arm using a $4f$ pulse shaper, consisting of two pairs of gratings (G) and cylindrical lenses (CL) and hosting a vortex plate (VP) in its Fourier plane. This method was used for the first experimental demonstrations of STOVs \cite{hancock_2019_free, chong_2020_generation} and remains the standard route to producing them. Available VPs allowed us to generate STOV pulses with topological charges (TCs) up to $\ell = \pm 3$. We have further described this shaper in \cite{gadeyne_2026_controlled}. With this arrangement, the pulse output by the shaper is not yet a STOV, it only forms at the focus of a subsequent lens (CL'), which may either be cylindrical (focusing along the $x$ axis only) or spherical (focusing also along $y$) -- differences between the two cases will be discussed in a later section.

To measure the space-time envelope of our pulses around this focal plane, we implemented spatially-resolved Fourier transform spectral interferometry (FTSI), a method first applied to focused STOVs in \cite{gui_2022_single} and illustrated in \autoref{fig1} (b-e). A very detailed account of our FTSI analysis procedure can be found in \cite{gadeyne_2026_spatiotemporal} and will not be restated here. Briefly, the principle of FTSI is to detect spectral fringes (along the angular frequency axis $\omega$) between a shaped pulse and a delayed reference pulse coming from the other arm of the interferometer (b). From them, the spatially resolved (along $x$) spectral phase difference between the shaped and reference pulses can be retrieved by Fourier-filtering techniques. To this end, we built an imaging spectrometer consisting of a grating (1200 lines/mm) followed by a cylindrical lens ($f=30$~cm) and CCD camera, theoretically offering a spectral resolution $\delta \omega \sim 1.6 \cdot 10^{-4}$ rad/fs (that is $\delta \lambda \sim 0.09$~nm). The spectral phase of the reference pulse was measured independently via frequency-resolved optical gating (FROG) \cite{trebino_2000_frequency} (see (c)), and was assumed to be spatially homogeneous (planar wavefronts). This remains a reasonable approximation even after extended free propagation of the reference beam, because the shaped pulse was observed at focus where it was always much smaller than the reference one. After subtraction of this reference phase, the intensity and phase of the shaped pulses are first obtained in the spatiospectral domain (d), then converted to the time domain (e) by a Fourier transform.

After collinear recombination of the two pulses, a translation stage was introduced to scan the distance $z$ from lens to detector over several dozens of centimeters, to reconstruct the pulse evolution across the entire focal region.
Finally, a variable free propagation distance could also be introduced after the shaper, amounting to a distance $D$ from the VP to the focusing lens CL'. The results section will discuss two different regimes with respect to this free propagation step.

\section{Analytical theory}

To model our experiment in an analytically-tractable way, we approximate the field in the $(X, \Omega = \omega - \omega_0)$ Fourier plane by a LG mode
\begin{align}
    \widetilde{\psi}(X, \Omega)
    =
    e^{-(\Omega/\Omega_0)^2}
    e^{-(X/X_0)^2}
    \left[ \frac{\Omega}{\Omega_0} \mp i \frac{X}{X_0}\right]^\aell
    ,
    \label{eq7:LG_shaper}
\end{align}
where leading normalization constants have been omitted, and $X_0$ and $\Omega_0$ are respectively the spatial and spectral widths of the pulse. 
In the time domain, the pulse envelope at this plane is the dual, THG mode \cite{porras_2025_optics}
\begin{align*}
    \widetilde{\psi}(X, t)
    & =
    \int \dd \Omega \,e^{-i\Omega t}
    \,
    \widetilde{\psi}(X, \Omega)
    \\
    & \propto
    e^{-(t/t_0)^2}
    e^{-(X/X_0)^2}
    \frac{1}{2^\aell}
    \mathcal{H}_\aell \left\{
    \frac{t}{t_0} \pm \frac{X}{X_0}\right\}
    ,
\end{align*}

where 
$t_0 = 2/\Omega_0$ is the temporal width, a factor $\Omega_0 \sqrt{\pi} (-i)^\aell$ has been omitted and $\mathcal{H}_\aell$ is a Hermite polynomial. \\

To model free propagation over a distance $D$, and subsequent propagation over $z$ after a lens of focal length $f$, two Fresnel diffraction integrals must be performed. These can be computed in a single step using the $\mathsf{ABCD}$ formalism put forward by Porras for Hermite/Laguerre-type STOVs in \cite{porras_2025_optics}. The relevant $\mathsf{ABCD}$ matrix is
\begin{align*}
    \begin{pmatrix}
        \mathsf{A} & \mathsf{B} \\
        \mathsf{C} & \mathsf{D} \\
    \end{pmatrix}
    & =
    \begin{pmatrix}
        1 & z \\
        0 & 1 \\
    \end{pmatrix}
    \begin{pmatrix}
        1 & 0 \\
        -1/f & 1 \\
    \end{pmatrix}
    \begin{pmatrix}
        1 & D \\
        0 & 1 \\
    \end{pmatrix}
    \\
    & =
    \begin{pmatrix}
        1-z/f
        & D \left(1-z/f\right) + z
        \\
        -1/f
        & 1 - D/f \\
    \end{pmatrix}
\end{align*}
and the propagated space-time envelope writes \cite{porras_2025_optics}
\begin{widetext}
\begin{align}
\begin{split}
    \psi(x, t ; z) 
    & =
    \frac{e^{-(t/t_0)^2}
    e^{i(Z_R/q_1)(x/X_0)^2}}
    {\sqrt{\mathsf{A} + \mathsf{B}/q_0}}
    \cdot
    \left(\frac{\mathsf{A}}{\mathsf{A}+\mathsf{B}/q_0}\right)^{\frac{\aell}{2}}
    \frac{1}{2^\aell}
    \mathcal{H}_\aell \left\{
    \left(\frac{\mathsf{A}}{\mathsf{A}+\mathsf{B}/q_0}\right)^{-\frac{1}{2}}
    \left[ 
    \frac{t}{t_0} \pm \frac{x}{X_0}  \frac{1}{\mathsf{A}+\mathsf{B}/q_0}
    \right]
    \right\}
    ,
\end{split}
\label{eq7:focused_stov}
\end{align}
\end{widetext}
where $Z_R = k_0 X_0^2/2$ is the Rayleigh length of the input beam, and complex beam parameters are
\begin{align*}
q_0 =-iZ_R
,
\qquad \qquad
q_1 = \frac{\mathsf{A}+\mathsf{B}/q_0}{\mathsf{C} + \mathsf{D}/q_0} 
.
\end{align*}
\\

\paragraph*{\bfseries Rescaling the model.}
This analytically-tractable model is only an approximation of our experiment, because the actual field in the Fourier plane of our shaper does not exactly take the LG form \eqref{eq7:LG_shaper}. Indeed, in a phase-only shaper the intensity distribution at the Fourier plane is that of the input beam (inset in \autoref{fig1} (a)), and is thus independent of the applied vortex phase. This $\ell$-independent spot size does not match the fact that the root-mean-squared radius of the LG modes \eqref{eq7:LG_shaper} grows with $\ell$ like $\sqrt{\aell + 1}$. As mentioned in \cite{gadeyne_2026_controlled}, the experimental STOV at focus is better described by hypergeometric-Gaussian (HyGG) modes \cite{karimi_2007_hypergeometric, bekshaev_2009_structure} -- but those are less amenable to analytical calculations. To improve the agreement between the LG/THG model and our experiment, we decided to rescale the analytical results to better match the actual divergence of the experimental beam. A natural way to quantify this divergence could have been through the well-known beam quality factor $M^2$, unfortunately this one cannot be defined in the usual fashion for HyGG modes as certain second-moment integrals do not converge \cite{vallone_2015_properties}. Instead, we simply decided to rescale the input LG mode \eqref{eq7:LG_shaper} by changing $X_0$ as a function of $\aell$, trying to implement a more constant spot size in the Fourier plane of our shaper. There is no unique way to do this -- for modeling the present experiments, we found good agreement by enforcing a constant root-mean-squared intensity radius in the Fourier plane, \emph{i.e.} using the rescaling
\begin{align}
    (X_0, \Omega_0) \to \frac{1}{\sqrt{\aell+ 1}}
    (X_0, \Omega_0).
    \label{eq:rescaling}
\end{align}
As a consequence, the Rayleigh length of the collimated beam gets effectively rescaled as $Z_R \to Z_R / (\aell+1)$. It should be emphasized that these rescalings are only heuristics, introduced to better align the analytically-tractable model with real-life results. \\

\paragraph*{\bfseries Coordinate of minimal radius.}
From \eqref{eq7:focused_stov}, the spatial Gaussian radius of the beam, $x'(z)$, can be put under the general form
\begin{align}
    {x'}^2(z)
    =
    \frac{X_0^2}{Z_R}
    \frac{1}{\Im{1/q_1}}
    =
    {x_0'}^2 \left[ 1 + \frac{(z-z_0)^2}{z_R^2} \right]
    \label{eq7:xp}
\end{align}
where $z_R = k_0 {x_0'}^2/2$ is the Rayleigh length of the focused beam, $x'_0$ its minimal radius, and $z_0$ the coordinate at which it is attained. Defining two dimensionless parameters,
\begin{align*}
    \alpha = \frac{D}{Z_R} 
    \qquad \textrm{and} \qquad
    \beta = \frac{f}{Z_R},
\end{align*}
some rearranging eventually leads to
\begin{align}
    z_0 = f\frac{1 - \alpha\beta + \alpha^2}{1 - 2\alpha\beta + \alpha^2 + \beta^2}.
    \label{eq7:z0_position}
\end{align}
The coordinate of minimal radius $z_0$ varies with $D$ and $f$. Note that even without propagation before the lens ($\alpha = 0$), we have $z_0 = f / (1+\beta^2) > f$: this effect is known as \emph{focal shift} \cite{self_1983_focusing}; because focal lengths typically used in experiment verify $f \ll Z_R$, this effect is rarely noticed. In contrast, free propagation can occur over several meters, which may well produce a sizable parameter $\alpha \gtrsim 0.1$.  \\

\begin{figure*}[t!]
    \centering
    \includegraphics[trim={0, 0pt, 0, 0}, width=\linewidth]{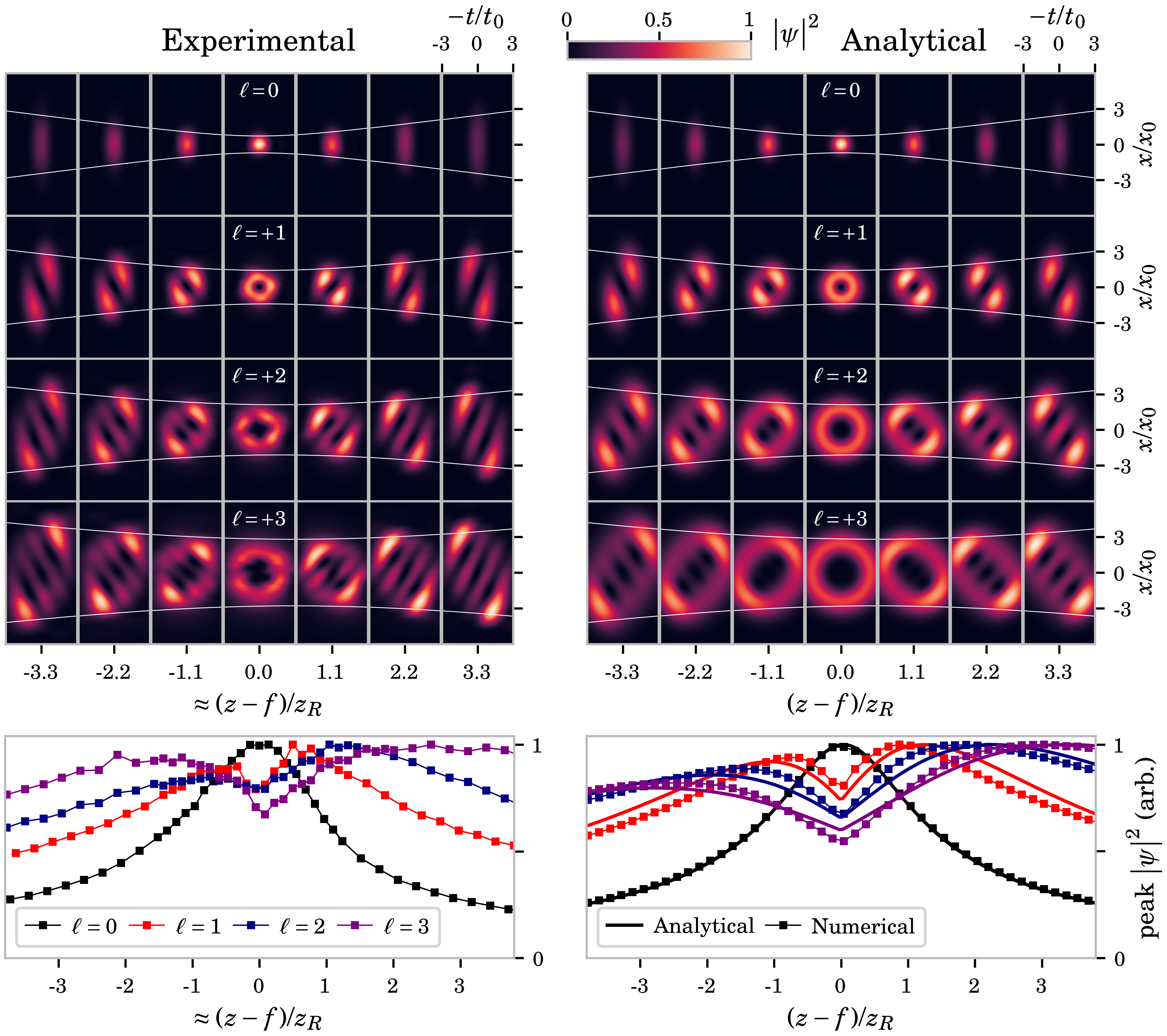}
    \caption{Top: experimental (left) and analytical (right) space-time intensity envelopes at several coordinates $z$ around the focus, after a moderate free propagation distance. Parameters are $D=1.5$~m, $f=1$~m, and the values $t_0=275$~fs and $x'_0 \approx x_0=154$~µm were retrieved from the experimental $\ell=0$ dataset ($z_R = k_0x_0^2/2 = 73$~mm, $X_0 = 2 f / k_0 x_0=2.1$~mm, $Z_R = k_0 X_0^2/2=13.8$~m). Different rows correspond to different $\ell$ and are normalized independently; in a given row all columns share the same normalization to highlight the evolution of peak intensity with propagation.  The white lines superimposed over each row show the analytical root-mean-square radius $x'(z) \sqrt{(\aell+1)/2}$ as a guide to the eye.
    \\
    Bottom: experimental (left) and theoretical (right) peak intensity in the space-time pulse as a function of the propagation distance, for the same conditions as the upper plots. Each curve is normalized to an overall maximal intensity of 1. The theory panel compares a numerical simulation of our pulse shaping setup to the analytical model \eqref{eq7:focused_stov}.
    }
    \label{fig2}
\end{figure*}

\paragraph*{\bfseries Coordinate of the circular STOV.}
Meanwhile, reshaping of the space-time mode is captured by the Hermite polynomial factor in \eqref{eq7:focused_stov}. Its argument is complex-valued, allowing this single factor to elegantly encode the full reshaping of the pulse from a tilted-Hermite to a Laguerre-type mode \cite{porras_2023_propagation, porras_2025_optics}. A Laguerre-type STOV with circular symmetry in the rescaled coordinates $(t/t_0, x/x')$ is obtained when $\mathsf{A} = 1-z/f = 0$. Indeed, all terms in the Hermite polynomial are then suppressed by the prefactor $(A/(A+B/q_0))^{\aell/2}$ except for the highest power, such that the second factor in \eqref{eq7:focused_stov} becomes
$
(-i)^\aell
\left[ 
t/t_0 \mp i x/x'(f)
\right]^\aell
$.
Therefore, the coordinate $z_\odot$ at which the circular STOV is realized is the geometrical focus,
\begin{align}
    z_\odot = f.
    \label{eq7:zdot_position}
\end{align}
In general, this coordinate will \emph{not} coincide with that of minimal radius, $z_0$.
\\

\paragraph*{\bfseries Dependence on the topological charge.}
Most importantly, the position \eqref{eq7:zdot_position} at which the circular STOV forms is always the geometrical focus, regardless of $\ell$. Although \eqref{eq7:z0_position} does not explicitly involve the topological charge $\ell$, it depends on it in practice through the effective Rayleigh length $Z_R \propto 1 / (\aell +1)$ mentioned earlier. In practice, $z_0$ gets displaced further downstream when increasing $\ell$.

\section{Results and discussion}

\subsection{Focusing after moderate free propagation}

We start with the traditional experimental configuration in which the pulse is focused shortly after exiting the shaper. We hence adopt the moderate free propagation distance $D \approx 1.5$~m.
The input beam for this experiment had a radius $X_0 \approx 2.1$~mm, corresponding to $Z_R \approx 13.8~$m. The cylindrical lens used to focus the STOV had a focal length $f=1$~m. This corresponds to parameters $\alpha \sim 0.11$ and $\beta \sim 0.07$. We measured the space-time pulse $\psi(x,t;z)$ for $\ell \in \{0, +1, +2, +3\}$ at $\sim 40$ coordinates $z$ spanning over several $z_R$. In \autoref{fig2}, we present the space-time pulse intensity at seven equidistant positions, forming a composite image of the pulse reshaping with propagation. Importantly, all envelopes are normalized to the same total energy, to render how intensity evolves when passing the focus.

With a moderate free propagation distance, the space-time mode of the STOV is found to evolve almost symmetrically on either side of the focus: a circular STOV is obtained near the position of minimal waist, and breaks into $\aell+1$ lobes with opposite slope on either side. This is the qualitative behaviour reported in previous experimental studies \cite{hancock_2019_free, hancock_2021_mode, gui_2022_single} and is what the analytical theory predicts (neglecting the focal shift) when no free propagation step is present \cite{porras_2024_procedure}. The right panel shows the analytical predictions of \eqref{eq7:focused_stov}: the evolution of the pulse shape is in good agreement, but the ring radius near $z=f$ is exaggerated. This is due to the aforementioned fact that our shaper does not produce LG but rather HyGG modes, and though we found the rescaling \eqref{eq:rescaling} to better reproduce properties of the pulse over several $z_R$, it does not quantitatively reproduce the ring radius at focus. This is likely due to the radial envelope of HyGG modes decaying much slower than that of LG modes at large radii. \\

\paragraph*{\bfseries Effect of modal reshaping on peak intensity.}
Inspection of \autoref{fig2} reveals an interesting behaviour for the \emph{peak intensity} with propagation: while for the $\ell=0$ pulse the peak intensity evidently coincides with the position of minimal waist, this is strikingly \emph{not} the case for $\ell \neq 0$ STOVs, for which noticeably higher intensities are reached at $z$ coordinates upstream and downstream of the focus, where the space-time mode is a hybrid of LG and THG functions. To emphasize this point, we plot in the lowermost panels of \autoref{fig2} the experimental and theoretical peak intensities as a function of $z$. Here, the theory panel includes both the rescaled analytical model \eqref{eq7:focused_stov} and a Fresnel-diffraction simulation of our shaper that accounts for the HyGG-type modal structure -- the two agree well over several $z_R$.

Unexpectedly, both theory and experiment reveal that the peak intensity of a STOV pulse reaches a local \emph{minimum} around its waist plane. This is because at this coordinate, the space-time intensity spreads evenly around a ring, leading to a lower peak value despite the tighter beam caustic. In parallel, local maxima are observed before and after the focus, as the pulse transitions into a THG-type mode in which intensity concentrates into two outer lobes, reaching higher peak values despite the overall divergence. Note that the slight asymmetry between the maxima before and after the waist plane is due to propagation distance $D$ being finite.

Strikingly, the longitudinal offset of these maxima with respect to the waist plane is quite significant, on the order of $z_R$ (or even more as $\ell$ increases). At this point, it should be emphasized that we have thus far considered focusing \emph{along the $x$ coordinate only} (with a cylindrical lens). In many applications however, a STOV would also be focused along $y$; and along that axis the pulse has a typically Gaussian profile -- we will return to this when illustrating consequences for harmonic generation.

\subsection{Focusing after extended free propagation}

\begin{figure*}[t!]
    \centering
    \includegraphics[trim={0, 0pt, 0, 0}, width=\linewidth]{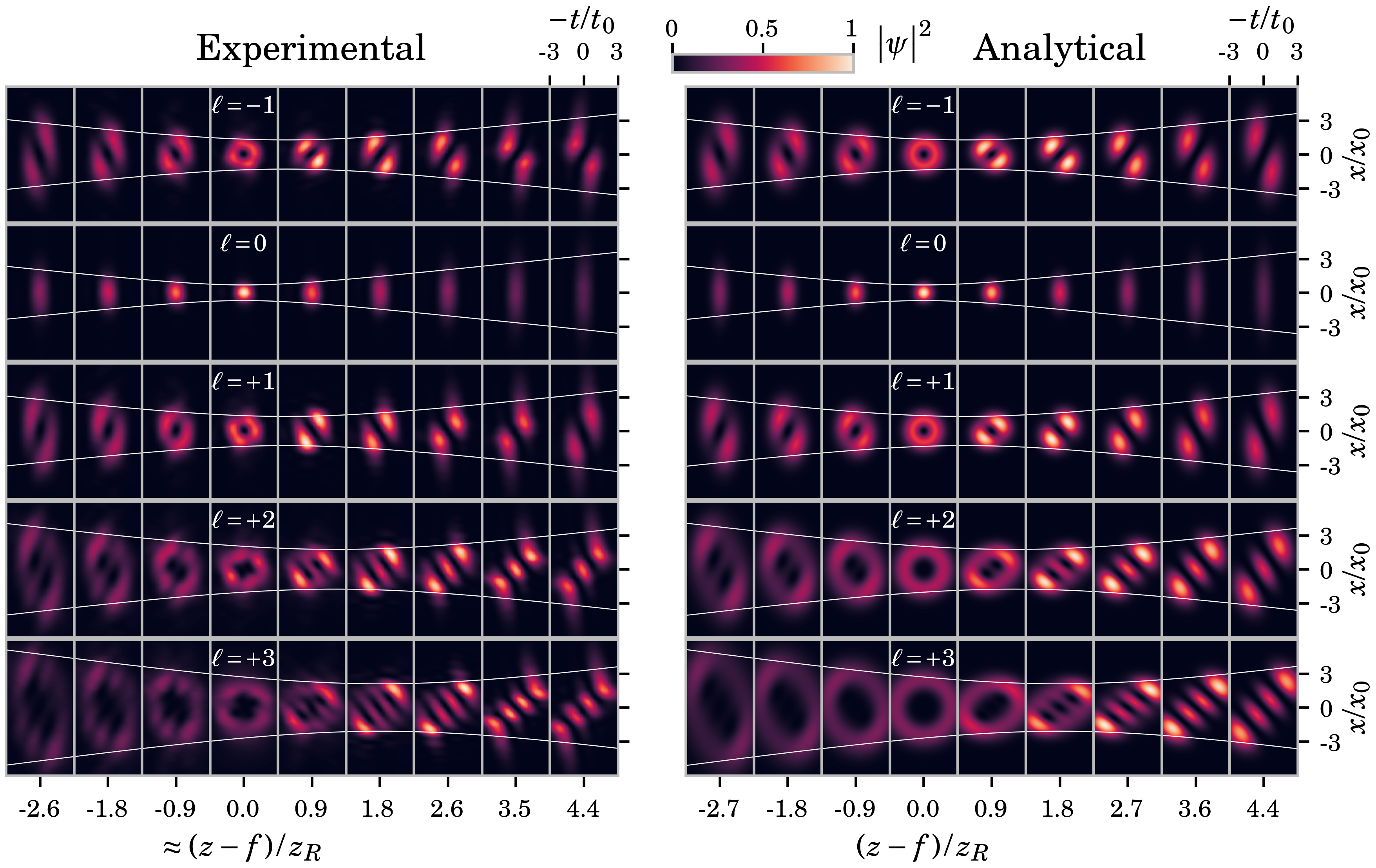}
    \caption{Experimental (left) and analytical (right) space-time intensity envelopes at several coordinates $z$ around the focus, after a long free propagation distance. Parameters are $D=6$~m, $f=1$~m, and the values $t_0=277$~fs and $x_0'\approx x_0=122$~µm were retrieved from the experimental $\ell=0$ dataset ($z_R = k_0x_0^2/2 = 45$~mm, $X_0 = 2 f / k_0 x_0=2.7$~mm, $Z_R=k_0X_0^2/2=22.0$~m). The presentation is analogous to \autoref{fig2}.
    }
    \label{fig3}
\end{figure*}

We now turn to the case of a long free propagation distance, $D = 6$~m, keeping the same $f=1$~m lens. For this experiment the input waist of the collimated beam was slightly different ($X_0=2.7$~mm), giving $\alpha \sim 0.27$,  $\beta \sim 0.04$. We measured the pulse $\psi(x,t;z)$ for $\ell \in \{-1, 0, +1, +2, +3\}$ at $\sim 30$ positions over several $z_R$, and present a selection in \autoref{fig3}.

As expected from the analytical model, a circular STOVs is always observed at the same coordinate $(z_\odot=f)$, irrespective of the topological charge $\ell$. However, the modal reshaping on either side of this position is now noticeably asymmetrical: while the mode downstream of the focus shows well-defined and intense lobes, the mode upstream looks more like a blurred ellipse. This asymmetry emerges because the position of minimal waist for the Gaussian caustic \eqref{eq7:z0_position} is pushed further downstream ($z_0 > f$) due to the long free propagation step $D$ before the lens. This effect is more visible for larger $\aell$, because the effective Rayleigh length of the collimated beam during this propagation step decreases like $1/(\aell + 1)$. Note that the asymmetry does not depend on the sign of the phase vortex: peak intensities are equivalently pushed downstream in both $\ell=+1$ and $\ell=-1$ datasets. As such, it does not seem appropriate to call this an effect of the transverse OAM of these pulses, but rather an effect of the broader angular spectrum introduced by vortex plates with larger $\aell$. Note that for the same reason, the Gaussian caustic of the more common \emph{spatial} LG vortex beams would also get pushed downstream in a similar manner when free propagation is introduced -- in this case however, the transverse spatial mode remains circularly-symmetric at any $z$, and the position of peak intensity would strictly coincide with that of minimal radius.

\subsection{Consequences for nonlinear optics}

The phenomena evidenced above may have particularly strong consequences when using STOV pulses to drive nonlinear optical processes, of which the efficiency varies sharply with intensity. One example is harmonic generation: to lowest order in perturbation theory (and assuming perfect phase-matching conditions), when driving the process with a space-time electric field $E(x,t) = \psi(x,t) e^{-i\omega_0 t}$ the field amplitude of harmonic $q$ will scale as $E^q$ \cite{reintjes_1984_nonlinear, boyd_nonlinear_2008}. This provides a way to transfer a STOV phase topology to shorter wavelengths, as demonstrated experimentally in \cite{gui_2021_second, hancock_2021_second, martinhernandez_2025_extreme, gadeyne_2026_controlled}. Due to the nonlinear amplitude scaling however, the process will be highly more efficient when placing the nonlinear medium at the coordinate where the highest peak intensity is reached -- and this is typically how its position would be selected experimentally. As shown in \autoref{fig3} however, this will not coincide with the position where the circular STOV is actually formed.

\begin{figure}[b!]
    \centering
    \includegraphics[width=\linewidth]{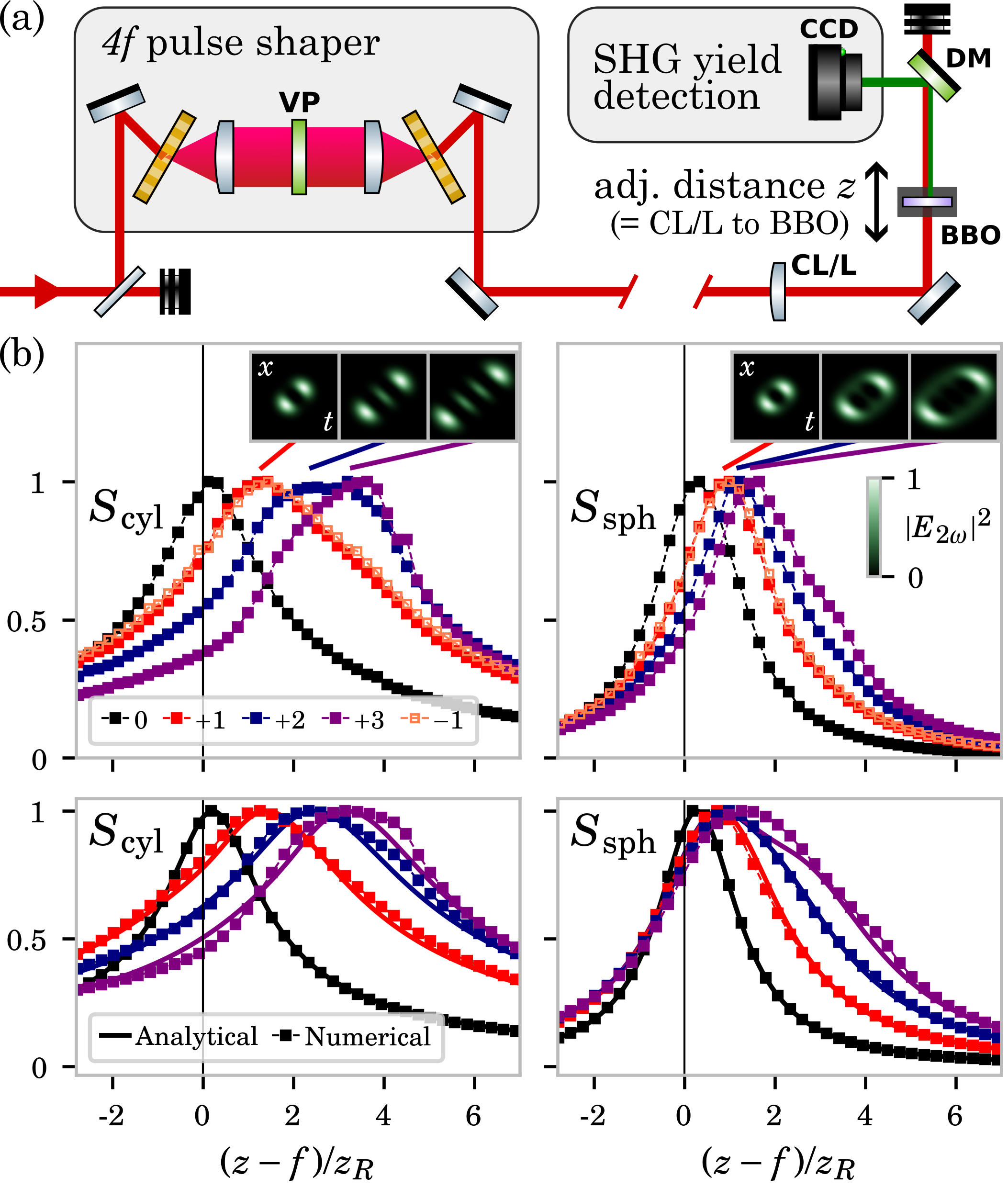}
    \caption{(a) Sketch of the optical setup for measuring the efficiency of STOV-driven SHG. The pulse is focused either with a cylindrical (CL) or spherical (L) lens. SHG occurs in a BBO crystal positioned at an adjustable distance $z$ from the lens. The $2\omega$ pulse is filtered from the fundamental by a dichroic mirror (DM) and collected by a CCD camera: the image is integrated to obtain the total SHG yield. (b) SHG yields, measured experimentally (upper panels) or theoretically (lower panels) either from a numerical simulation of our setup or from the rescaled analytical model. Yield curves in the left/right columns correspond to cylindrical/spherical focusing of the STOV pulse.  All curves are independently normalized to a maximum value of 1. Insets show the analytical $2\omega$ space-time intensity at the exit of the crystal for the $z$ at which the maximal yield was observed in experiment.
    }
    \label{fig4}
\end{figure}

To illustrate this effect, we modified our setup as shown in \autoref{fig4} (a). We drove second-harmonic generation (SHG, $q=2$) in a baryum borate crystal (BBO type I, $1$-mm thick). The crystal was mounted on a translation stage to scan its position $z$ with respect to the focusing lens. The 515-nm field was filtered from the remaining fundamental with a dichroic mirror (DM) and directed to a CCD camera. The total SHG yield was obtained by integrating the collected intensity. Assuming perfect phase-matching and disregarding the thickness of the crystal, the $2\omega$ field at its exit can be simply modeled by
\begin{align*}
    E_{2\omega}(x, y, t ; z) 
    & \propto E^2(x,y,t ; z) 
    \\
    & \propto \big[\psi(x,t ; z) \cdot f(y ; z)\big]^2 e^{-i2\omega t}
\end{align*}
where $f(y)$ is the envelope of the pulse along the $y$ axis. We considered two situations, (i) cylindrical focusing along $x$ only, as in the FTSI experiments of \autoref{fig2} and \autoref{fig3} and (ii) spherical focusing (along $x$ and $y$), as would typically be done in a harmonic generation experiment. In the first case the envelope can be considered Gaussian and roughly constant over the scanned distance, $f(y) \propto e^{-(y/Y_0)^2}$, where $Y_0$ is the waist of the collimated beam along the horizontal axis (on our setup, the beam was slightly elliptical and from a fit of the $l=0$ caustic we estimated $Y_0 \approx 0.85 X_0$). In this case the SHG yield evolves with $z$ according to
\begin{align*}
    S_\textrm{cyl}(z)
    & \propto
    \iiint \dd x \dd y \dd t ~
    \abs{E_{2\omega}(x, y, t ; z)}^2
    \\
    & \propto
    \iint \dd x \dd t ~
    \abs{\psi(x, t ; z)}^4
    .
\end{align*}

In the case of spherical focusing, the envelope along $y$ also varies with $z$, and its evolution can be obtained from \eqref{eq7:focused_stov} with $\ell=0$, $f(y; z) \propto e^{-[y/y'(z)]^2} /\sqrt{y'(z)}$ with
\begin{align*}
    y'^2(z)
    =
    Y_0^2
    \abs{\mathsf{A}+ i \mathsf{B}/(k_0 Y_0^2/2)}^2
    .
\end{align*}
SHG yields then vary with $z$ as
\begin{align*}
    S_\textrm{sph}(z)
    & \propto
    S_\textrm{cyl}(z)
    \cdot
    \int \dd y ~
    f(y ; z)
    \\
    & \propto
    S_\textrm{cyl}(z)
    \cdot
    \frac{1}{y'(z)}
    .
\end{align*} \\

\paragraph*{\bfseries Cylindrical focusing.}
In the top left panel of \autoref{fig4} (b), we plot experimental SHG yields obtained with $D=6$~m free propagation before cylindrical focusing. In accordance with the peak intensities of \autoref{fig3}, the maximal yield is clearly displaced downstream of $z=f$ as $\aell$ increases, by up to several $z_R$. In the bottom panel, these curves are compared to a numerical simulation (squares) taking into account parameters from our pulse shaper, and to the above analytical results (lines). Here again theory and experiment agree convincingly. Note that the experimental yield from the $\ell = -1$ STOV overlaps almost perfectly with the $\ell=+1$ curve, in accordance with a dependence of the intensity on $\aell$ only.

In the insets, we show the analytical space-time intensity of the $2\omega$ pulse generated at the exit of the crystal, assuming the latter to be positioned at a $z$ around which the highest yield was observed in experiment. Here, the $2\omega$ pulse is not expected to be round: instead, it looks like a THG mode elevated to its second power, in agreement with the space-time mode of the driving pulse measured at the corresponding $z$ in \autoref{fig3}. \\

\paragraph*{\bfseries Spherical focusing.}
In a conventional harmonic generation experiment however, the driver would actually be focused with a spherical lens, to maximize its intensity. In the top right panel of \autoref{fig4} (b), we show the experimental yields measured with spherical focusing. Here, the waist along $y$ is always minimized close to $z=f$, regardless of $\ell$. The focused Rayleigh length along $y$ is also shorter than that along $x$ for $\aell \neq 0$. As a consequence, the yield curves are ``pinched'' more sharply along $z$, and the maximal yield coordinates are all shifted closer to $z=f$. In these conditions, the trend displayed by the theory is matching fairly well the experiment. We note however  that the yield peaks look slightly too broad and not shifted enough with respect to each other, in spite of taking into account the ellipticity of the input beam, and checking that the $x$ and $y$ foci occurred at the same $z$. We have to attribute these discrepancies to further defects of our beam (non-Gaussian shape and/or $M^2>1$, clipping...) or to unnoticed phase-matching or alignment issues with the BBO.

Still, even with spherical focusing there remains a noticeable downstream shift of the maximal yield, of one to two Rayleigh lengths, and the analytical $2\omega$ pulse (insets) generated at this position already looks like a tilted ellipse, a hybrid of THG and LG modes. These observations lead us to conclude that to properly drive a nonlinear process with a circularly-symmetric STOV pulse, the placement of the nonlinear medium should not be made by maximizing the yield. This is particularly true if considerable free propagation occurs between the shaping and focusing stages of the beamline. If this was done, the resulting pulse would not propagate like an ideal, circularly-symmetric STOV. Moreover, its intrinsic transverse OAM would likely be different: a recent numerical study of STOV-driven high-harmonic generation showed that the OAM of the harmonic pulse depends critically on the relative $z$ position of the waist plane and generating medium \cite{martinhernandez_2025_isolateda}. Finally, we note that these effects generalize beyond STOVs, to any complex space-time pulse -- they would only become stronger when focusing a three-dimensional structured pulse (such as \eg toroidal vortices \cite{wan_2022_toroidal} or vortex knots and links \cite{zhou_2026_spatiotemporally}), in which case the embedding caustic along $y$ could also participate in the downstream shift of the peak intensity.

\section{Conclusions}

In summary, we have studied analytically and experimentally the focusing of STOV pulses produced by a $4f$ pulse shaper, to describe the evolution of their intensity around the focal plane in a quantitative manner.
We have unveiled a non-trivial dependence of their peak intensity: first, their inherent reshaping between LG- and THG-type space-time modes displaces the coordinate at which the energy is most concentrated within the pulse away from where the STOV is circular; second, introducing substantial free propagation prior to focusing the pulse leads to a longitudinal shift between the coordinate at which the circular STOV is formed and that at which the Gaussian caustic is tightest. These observations have direct consequences for the upconversion of STOV pulses (and other complex space-time pulses) in nonlinear optics; as an example we have reported how the yield of second-harmonic generation could be maximized several Rayleigh lengths downstream of the coordinate at which a circular STOV forms. Our study signals careful characterization of space-time pulses and of their propagation as a requisite for their application to nonlinear optics and strong-field physics. As a side benefit, this study also highlights the practical differences between analytical models based on Hermite- and Laguerre-Gaussian modes, and real-life light fields.

\section*{Acknowledgments}

This work was supported by state funding within the PEPR LUMA, managed by the National Research Agency under the France 2030 investment plan bearing the reference ANR-22-EXLU-0002 and the moonshot project ``Tornado"; it received funding from the Agence Nationale pour la Recherche (contract No. ANR HELIMAG ANR-21-CE30-0037), the Indo-French CEFIPRA Grant Project No.- (7104-I), and from the European Union’s Horizon Europe research and innovation programme under grant agreement No 101129641 (OBELIX).

\section*{Author contributions}

T.G. performed the experiments, analyzed the data, developed the analytical and numerical theory and wrote the first draft of the manuscript. V.V. helped with the experiments. T.R. conceptualized the work, provided guidance on all parts of the project and contributed to the manuscript.

\bibliography{allV2}

\begin{thebibliography}{35}%
\makeatletter
\providecommand \@ifxundefined [1]{%
 \@ifx{#1\undefined}
}%
\providecommand \@ifnum [1]{%
 \ifnum #1\expandafter \@firstoftwo
 \else \expandafter \@secondoftwo
 \fi
}%
\providecommand \@ifx [1]{%
 \ifx #1\expandafter \@firstoftwo
 \else \expandafter \@secondoftwo
 \fi
}%
\providecommand \natexlab [1]{#1}%
\providecommand \enquote  [1]{``#1''}%
\providecommand \bibnamefont  [1]{#1}%
\providecommand \bibfnamefont [1]{#1}%
\providecommand \citenamefont [1]{#1}%
\providecommand \href@noop [0]{\@secondoftwo}%
\providecommand \href [0]{\begingroup \@sanitize@url \@href}%
\providecommand \@href[1]{\@@startlink{#1}\@@href}%
\providecommand \@@href[1]{\endgroup#1\@@endlink}%
\providecommand \@sanitize@url [0]{\catcode `\\12\catcode `\$12\catcode `\&12\catcode `\#12\catcode `\^12\catcode `\_12\catcode `\%12\relax}%
\providecommand \@@startlink[1]{}%
\providecommand \@@endlink[0]{}%
\providecommand \url  [0]{\begingroup\@sanitize@url \@url }%
\providecommand \@url [1]{\endgroup\@href {#1}{\urlprefix }}%
\providecommand \urlprefix  [0]{URL }%
\providecommand \Eprint [0]{\href }%
\providecommand \doibase [0]{https://doi.org/}%
\providecommand \selectlanguage [0]{\@gobble}%
\providecommand \bibinfo  [0]{\@secondoftwo}%
\providecommand \bibfield  [0]{\@secondoftwo}%
\providecommand \translation [1]{[#1]}%
\providecommand \BibitemOpen [0]{}%
\providecommand \bibitemStop [0]{}%
\providecommand \bibitemNoStop [0]{.\EOS\space}%
\providecommand \EOS [0]{\spacefactor3000\relax}%
\providecommand \BibitemShut  [1]{\csname bibitem#1\endcsname}%
\let\auto@bib@innerbib\@empty
\bibitem [{\citenamefont {Hancock}\ \emph {et~al.}(2019)\citenamefont {Hancock}, \citenamefont {Zahedpour}, \citenamefont {Goffin},\ and\ \citenamefont {Milchberg}}]{hancock_2019_free}%
  \BibitemOpen
  \bibfield  {author} {\bibinfo {author} {\bibfnamefont {S.~W.}\ \bibnamefont {Hancock}}, \bibinfo {author} {\bibfnamefont {S.}~\bibnamefont {Zahedpour}}, \bibinfo {author} {\bibfnamefont {A.}~\bibnamefont {Goffin}},\ and\ \bibinfo {author} {\bibfnamefont {H.~M.}\ \bibnamefont {Milchberg}},\ }\bibfield  {title} {\bibinfo {title} {Free-space propagation of spatiotemporal optical vortices},\ }\href {https://doi.org/10.1364/optica.6.001547} {\bibfield  {journal} {\bibinfo  {journal} {Optica}\ }\textbf {\bibinfo {volume} {6}},\ \bibinfo {pages} {1547} (\bibinfo {year} {2019})}\BibitemShut {NoStop}%
\bibitem [{\citenamefont {Chong}\ \emph {et~al.}(2020)\citenamefont {Chong}, \citenamefont {Wan}, \citenamefont {Chen},\ and\ \citenamefont {Zhan}}]{chong_2020_generation}%
  \BibitemOpen
  \bibfield  {author} {\bibinfo {author} {\bibfnamefont {A.}~\bibnamefont {Chong}}, \bibinfo {author} {\bibfnamefont {C.}~\bibnamefont {Wan}}, \bibinfo {author} {\bibfnamefont {J.}~\bibnamefont {Chen}},\ and\ \bibinfo {author} {\bibfnamefont {Q.}~\bibnamefont {Zhan}},\ }\bibfield  {title} {\bibinfo {title} {Generation of spatiotemporal optical vortices with controllable transverse orbital angular momentum},\ }\href {https://doi.org/10.1038/s41566-020-0587-z} {\bibfield  {journal} {\bibinfo  {journal} {Nature Photonics}\ }\textbf {\bibinfo {volume} {14}},\ \bibinfo {pages} {350} (\bibinfo {year} {2020})}\BibitemShut {NoStop}%
\bibitem [{\citenamefont {Allen}\ \emph {et~al.}(1992)\citenamefont {Allen}, \citenamefont {Beijersbergen}, \citenamefont {Spreeuw},\ and\ \citenamefont {Woerdman}}]{allen_1992_orbital}%
  \BibitemOpen
  \bibfield  {author} {\bibinfo {author} {\bibfnamefont {L.}~\bibnamefont {Allen}}, \bibinfo {author} {\bibfnamefont {M.~W.}\ \bibnamefont {Beijersbergen}}, \bibinfo {author} {\bibfnamefont {R.~J.~C.}\ \bibnamefont {Spreeuw}},\ and\ \bibinfo {author} {\bibfnamefont {J.~P.}\ \bibnamefont {Woerdman}},\ }\bibfield  {title} {\bibinfo {title} {Orbital angular momentum of light and the transformation of {Laguerre}-{Gaussian} laser modes},\ }\href {https://doi.org/10.1103/PhysRevA.45.8185} {\bibfield  {journal} {\bibinfo  {journal} {Physical Review A}\ }\textbf {\bibinfo {volume} {45}},\ \bibinfo {pages} {8185} (\bibinfo {year} {1992})}\BibitemShut {NoStop}%
\bibitem [{\citenamefont {Hancock}\ \emph {et~al.}(2024)\citenamefont {Hancock}, \citenamefont {Zahedpour}, \citenamefont {Goffin},\ and\ \citenamefont {Milchberg}}]{hancock_2024_spatiotemporal}%
  \BibitemOpen
  \bibfield  {author} {\bibinfo {author} {\bibfnamefont {S.}~\bibnamefont {Hancock}}, \bibinfo {author} {\bibfnamefont {S.}~\bibnamefont {Zahedpour}}, \bibinfo {author} {\bibfnamefont {A.}~\bibnamefont {Goffin}},\ and\ \bibinfo {author} {\bibfnamefont {H.}~\bibnamefont {Milchberg}},\ }\bibfield  {title} {\bibinfo {title} {Spatiotemporal torquing of light},\ }\href {https://doi.org/10.1103/physrevx.14.011031} {\bibfield  {journal} {\bibinfo  {journal} {Physical Review X}\ }\textbf {\bibinfo {volume} {14}},\ \bibinfo {pages} {011031} (\bibinfo {year} {2024})}\BibitemShut {NoStop}%
\bibitem [{\citenamefont {Hancock}\ \emph {et~al.}(2025)\citenamefont {Hancock}, \citenamefont {Tripathi}, \citenamefont {Le}, \citenamefont {Goffin},\ and\ \citenamefont {Milchberg}}]{hancock_2025_transverse}%
  \BibitemOpen
  \bibfield  {author} {\bibinfo {author} {\bibfnamefont {S.~W.}\ \bibnamefont {Hancock}}, \bibinfo {author} {\bibfnamefont {N.}~\bibnamefont {Tripathi}}, \bibinfo {author} {\bibfnamefont {M.~S.}\ \bibnamefont {Le}}, \bibinfo {author} {\bibfnamefont {A.}~\bibnamefont {Goffin}},\ and\ \bibinfo {author} {\bibfnamefont {H.~M.}\ \bibnamefont {Milchberg}},\ }\bibfield  {title} {\bibinfo {title} {Transverse orbital angular momentum of amplitude perturbed fields},\ }\href {https://doi.org/10.1515/nanoph-2024-0595} {\bibfield  {journal} {\bibinfo  {journal} {Nanophotonics}\ }\textbf {\bibinfo {volume} {14}},\ \bibinfo {pages} {777} (\bibinfo {year} {2025})}\BibitemShut {NoStop}%
\bibitem [{\citenamefont {Gui}\ \emph {et~al.}(2021)\citenamefont {Gui}, \citenamefont {Brooks}, \citenamefont {Kapteyn}, \citenamefont {Murnane},\ and\ \citenamefont {Liao}}]{gui_2021_second}%
  \BibitemOpen
  \bibfield  {author} {\bibinfo {author} {\bibfnamefont {G.}~\bibnamefont {Gui}}, \bibinfo {author} {\bibfnamefont {N.~J.}\ \bibnamefont {Brooks}}, \bibinfo {author} {\bibfnamefont {H.~C.}\ \bibnamefont {Kapteyn}}, \bibinfo {author} {\bibfnamefont {M.~M.}\ \bibnamefont {Murnane}},\ and\ \bibinfo {author} {\bibfnamefont {C.-T.}\ \bibnamefont {Liao}},\ }\bibfield  {title} {\bibinfo {title} {Second-harmonic generation and the conservation of spatiotemporal orbital angular momentum of light},\ }\href {https://doi.org/10.1038/s41566-021-00841-8} {\bibfield  {journal} {\bibinfo  {journal} {Nature Photonics}\ }\textbf {\bibinfo {volume} {15}},\ \bibinfo {pages} {608} (\bibinfo {year} {2021})}\BibitemShut {NoStop}%
\bibitem [{\citenamefont {Hancock}\ \emph {et~al.}(2021{\natexlab{a}})\citenamefont {Hancock}, \citenamefont {Zahedpour},\ and\ \citenamefont {Milchberg}}]{hancock_2021_second}%
  \BibitemOpen
  \bibfield  {author} {\bibinfo {author} {\bibfnamefont {S.~W.}\ \bibnamefont {Hancock}}, \bibinfo {author} {\bibfnamefont {S.}~\bibnamefont {Zahedpour}},\ and\ \bibinfo {author} {\bibfnamefont {H.~M.}\ \bibnamefont {Milchberg}},\ }\bibfield  {title} {\bibinfo {title} {Second-harmonic generation of spatiotemporal optical vortices and conservation of orbital angular momentum},\ }\href {https://doi.org/10.1364/optica.422743} {\bibfield  {journal} {\bibinfo  {journal} {Optica}\ }\textbf {\bibinfo {volume} {8}},\ \bibinfo {pages} {594} (\bibinfo {year} {2021}{\natexlab{a}})}\BibitemShut {NoStop}%
\bibitem [{\citenamefont {Fang}\ \emph {et~al.}(2021)\citenamefont {Fang}, \citenamefont {Lu},\ and\ \citenamefont {Liu}}]{fang_2021_controlling}%
  \BibitemOpen
  \bibfield  {author} {\bibinfo {author} {\bibfnamefont {Y.}~\bibnamefont {Fang}}, \bibinfo {author} {\bibfnamefont {S.}~\bibnamefont {Lu}},\ and\ \bibinfo {author} {\bibfnamefont {Y.}~\bibnamefont {Liu}},\ }\bibfield  {title} {\bibinfo {title} {Controlling photon transverse orbital angular momentum in high harmonic generation},\ }\href {https://doi.org/10.1103/physrevlett.127.273901} {\bibfield  {journal} {\bibinfo  {journal} {Physical Review Letters}\ }\textbf {\bibinfo {volume} {127}},\ \bibinfo {pages} {273901} (\bibinfo {year} {2021})}\BibitemShut {NoStop}%
\bibitem [{\citenamefont {Martín-Hernández}\ \emph {et~al.}(2025{\natexlab{a}})\citenamefont {Martín-Hernández}, \citenamefont {Gui}, \citenamefont {Plaja}, \citenamefont {Kapteyn}, \citenamefont {Murnane}, \citenamefont {Liao}, \citenamefont {Porras},\ and\ \citenamefont {Hernández-García}}]{martinhernandez_2025_extreme}%
  \BibitemOpen
  \bibfield  {author} {\bibinfo {author} {\bibfnamefont {R.}~\bibnamefont {Martín-Hernández}}, \bibinfo {author} {\bibfnamefont {G.}~\bibnamefont {Gui}}, \bibinfo {author} {\bibfnamefont {L.}~\bibnamefont {Plaja}}, \bibinfo {author} {\bibfnamefont {H.~C.}\ \bibnamefont {Kapteyn}}, \bibinfo {author} {\bibfnamefont {M.~M.}\ \bibnamefont {Murnane}}, \bibinfo {author} {\bibfnamefont {C.-T.}\ \bibnamefont {Liao}}, \bibinfo {author} {\bibfnamefont {M.~A.}\ \bibnamefont {Porras}},\ and\ \bibinfo {author} {\bibfnamefont {C.}~\bibnamefont {Hernández-García}},\ }\bibfield  {title} {\bibinfo {title} {Extreme-ultraviolet spatiotemporal vortices via high harmonic generation},\ }\href {https://doi.org/10.1038/s41566-025-01699-w} {\bibfield  {journal} {\bibinfo  {journal} {Nature Photonics}\ }\textbf {\bibinfo {volume} {19}},\ \bibinfo {pages} {817} (\bibinfo {year} {2025}{\natexlab{a}})}\BibitemShut {NoStop}%
\bibitem [{\citenamefont {Gadeyne}\ \emph {et~al.}(2026)\citenamefont {Gadeyne}, \citenamefont {Vishnoi},\ and\ \citenamefont {Ruchon}}]{gadeyne_2026_controlled}%
  \BibitemOpen
  \bibfield  {author} {\bibinfo {author} {\bibfnamefont {T.}~\bibnamefont {Gadeyne}}, \bibinfo {author} {\bibfnamefont {V.}~\bibnamefont {Vishnoi}},\ and\ \bibinfo {author} {\bibfnamefont {T.}~\bibnamefont {Ruchon}},\ }\href {https://doi.org/10.48550/ARXIV.2608.00233} {\bibinfo {title} {Controlled generation of high-harmonic spatiotemporal optical vortices}} (\bibinfo {year} {2026})\BibitemShut {NoStop}%
\bibitem [{\citenamefont {Bliokh}(2021)}]{bliokh_2021_spatiotemporal}%
  \BibitemOpen
  \bibfield  {author} {\bibinfo {author} {\bibfnamefont {K.~Y.}\ \bibnamefont {Bliokh}},\ }\bibfield  {title} {\bibinfo {title} {Spatiotemporal vortex pulses: Angular momenta and spin-orbit interaction},\ }\href {https://doi.org/10.1103/physrevlett.126.243601} {\bibfield  {journal} {\bibinfo  {journal} {Physical Review Letters}\ }\textbf {\bibinfo {volume} {126}},\ \bibinfo {pages} {243601} (\bibinfo {year} {2021})}\BibitemShut {NoStop}%
\bibitem [{\citenamefont {Mazanov}\ \emph {et~al.}(2022)\citenamefont {Mazanov}, \citenamefont {Sugic}, \citenamefont {Alonso}, \citenamefont {Nori},\ and\ \citenamefont {Bliokh}}]{mazanov_2022_transverse}%
  \BibitemOpen
  \bibfield  {author} {\bibinfo {author} {\bibfnamefont {M.}~\bibnamefont {Mazanov}}, \bibinfo {author} {\bibfnamefont {D.}~\bibnamefont {Sugic}}, \bibinfo {author} {\bibfnamefont {M.~A.}\ \bibnamefont {Alonso}}, \bibinfo {author} {\bibfnamefont {F.}~\bibnamefont {Nori}},\ and\ \bibinfo {author} {\bibfnamefont {K.~Y.}\ \bibnamefont {Bliokh}},\ }\bibfield  {title} {\bibinfo {title} {Transverse shifts and time delays of spatiotemporal vortex pulses reflected and refracted at a planar interface},\ }\href {https://doi.org/10.1515/nanoph-2021-0294} {\bibfield  {journal} {\bibinfo  {journal} {Nanophotonics}\ }\textbf {\bibinfo {volume} {11}},\ \bibinfo {pages} {737} (\bibinfo {year} {2022})}\BibitemShut {NoStop}%
\bibitem [{\citenamefont {Zhang}\ \emph {et~al.}(2022)\citenamefont {Zhang}, \citenamefont {Ji},\ and\ \citenamefont {Shen}}]{zhang_2022_intense}%
  \BibitemOpen
  \bibfield  {author} {\bibinfo {author} {\bibfnamefont {L.}~\bibnamefont {Zhang}}, \bibinfo {author} {\bibfnamefont {L.}~\bibnamefont {Ji}},\ and\ \bibinfo {author} {\bibfnamefont {B.}~\bibnamefont {Shen}},\ }\bibfield  {title} {\bibinfo {title} {Intense harmonic generation driven by a relativistic spatiotemporal vortex beam},\ }\href {https://doi.org/10.1017/hpl.2022.38} {\bibfield  {journal} {\bibinfo  {journal} {High Power Laser Science and Engineering}\ }\textbf {\bibinfo {volume} {10}},\ \bibinfo {pages} {e46} (\bibinfo {year} {2022})}\BibitemShut {NoStop}%
\bibitem [{\citenamefont {Chen}\ \emph {et~al.}(2022)\citenamefont {Chen}, \citenamefont {Hu}, \citenamefont {Zhang},\ and\ \citenamefont {Yuan}}]{chen_2022_relativistic}%
  \BibitemOpen
  \bibfield  {author} {\bibinfo {author} {\bibfnamefont {Z.-Y.}\ \bibnamefont {Chen}}, \bibinfo {author} {\bibfnamefont {R.}~\bibnamefont {Hu}}, \bibinfo {author} {\bibfnamefont {S.}~\bibnamefont {Zhang}},\ and\ \bibinfo {author} {\bibfnamefont {T.}~\bibnamefont {Yuan}},\ }\bibfield  {title} {\bibinfo {title} {Relativistic high-order harmonic generation of spatiotemporal optical vortices},\ }\href {https://doi.org/10.1103/PhysRevA.106.013516} {\bibfield  {journal} {\bibinfo  {journal} {Physical Review A}\ }\textbf {\bibinfo {volume} {106}},\ \bibinfo {pages} {013516} (\bibinfo {year} {2022})}\BibitemShut {NoStop}%
\bibitem [{\citenamefont {Chen}\ \emph {et~al.}(2023)\citenamefont {Chen}, \citenamefont {Zhou}, \citenamefont {Li}, \citenamefont {Liu}, \citenamefont {Ciappina},\ and\ \citenamefont {Lu}}]{chen_2023_atomic}%
  \BibitemOpen
  \bibfield  {author} {\bibinfo {author} {\bibfnamefont {Y.}~\bibnamefont {Chen}}, \bibinfo {author} {\bibfnamefont {Y.}~\bibnamefont {Zhou}}, \bibinfo {author} {\bibfnamefont {M.}~\bibnamefont {Li}}, \bibinfo {author} {\bibfnamefont {K.}~\bibnamefont {Liu}}, \bibinfo {author} {\bibfnamefont {M.~F.}\ \bibnamefont {Ciappina}},\ and\ \bibinfo {author} {\bibfnamefont {P.}~\bibnamefont {Lu}},\ }\bibfield  {title} {\bibinfo {title} {Atomic photoionization by spatiotemporal optical vortex pulses},\ }\href {https://doi.org/10.1103/physreva.107.033112} {\bibfield  {journal} {\bibinfo  {journal} {Physical Review A}\ }\textbf {\bibinfo {volume} {107}},\ \bibinfo {pages} {033112} (\bibinfo {year} {2023})}\BibitemShut {NoStop}%
\bibitem [{\citenamefont {Rodríguez-Fortuño}\ \emph {et~al.}(2025)\citenamefont {Rodríguez-Fortuño}, \citenamefont {Picardi},\ and\ \citenamefont {Bliokh}}]{rodriguezfortuno_2025_bulk}%
  \BibitemOpen
  \bibfield  {author} {\bibinfo {author} {\bibfnamefont {F.~J.}\ \bibnamefont {Rodríguez-Fortuño}}, \bibinfo {author} {\bibfnamefont {M.~F.}\ \bibnamefont {Picardi}},\ and\ \bibinfo {author} {\bibfnamefont {K.~Y.}\ \bibnamefont {Bliokh}},\ }\bibfield  {title} {\bibinfo {title} {Bulk-to-surface coupling of spatiotemporal vortices},\ }\href {https://doi.org/10.1364/ol.572778} {\bibfield  {journal} {\bibinfo  {journal} {Optics Letters}\ }\textbf {\bibinfo {volume} {50}},\ \bibinfo {pages} {5534} (\bibinfo {year} {2025})}\BibitemShut {NoStop}%
\bibitem [{\citenamefont {Hancock}\ \emph {et~al.}(2021{\natexlab{b}})\citenamefont {Hancock}, \citenamefont {Zahedpour},\ and\ \citenamefont {Milchberg}}]{hancock_2021_mode}%
  \BibitemOpen
  \bibfield  {author} {\bibinfo {author} {\bibfnamefont {S.}~\bibnamefont {Hancock}}, \bibinfo {author} {\bibfnamefont {S.}~\bibnamefont {Zahedpour}},\ and\ \bibinfo {author} {\bibfnamefont {H.}~\bibnamefont {Milchberg}},\ }\bibfield  {title} {\bibinfo {title} {Mode structure and orbital angular momentum of spatiotemporal optical vortex pulses},\ }\href {https://doi.org/10.1103/physrevlett.127.193901} {\bibfield  {journal} {\bibinfo  {journal} {Physical Review Letters}\ }\textbf {\bibinfo {volume} {127}},\ \bibinfo {pages} {193901} (\bibinfo {year} {2021}{\natexlab{b}})}\BibitemShut {NoStop}%
\bibitem [{\citenamefont {Bliokh}(2023)}]{bliokh_2023_orbital}%
  \BibitemOpen
  \bibfield  {author} {\bibinfo {author} {\bibfnamefont {K.~Y.}\ \bibnamefont {Bliokh}},\ }\bibfield  {title} {\bibinfo {title} {Orbital angular momentum of optical, acoustic, and quantum-mechanical spatiotemporal vortex pulses},\ }\href {https://doi.org/10.1103/physreva.107.l031501} {\bibfield  {journal} {\bibinfo  {journal} {Physical Review A}\ }\textbf {\bibinfo {volume} {107}},\ \bibinfo {pages} {l031501} (\bibinfo {year} {2023})}\BibitemShut {NoStop}%
\bibitem [{\citenamefont {Porras}(2024)}]{porras_2024_clarification}%
  \BibitemOpen
  \bibfield  {author} {\bibinfo {author} {\bibfnamefont {M.~A.}\ \bibnamefont {Porras}},\ }\bibfield  {title} {\bibinfo {title} {Clarification of the transverse orbital angular momentum of spatiotemporal optical vortices},\ }\href {https://doi.org/10.1088/2040-8986/ad645a} {\bibfield  {journal} {\bibinfo  {journal} {Journal of Optics}\ }\textbf {\bibinfo {volume} {26}},\ \bibinfo {pages} {095601} (\bibinfo {year} {2024})}\BibitemShut {NoStop}%
\bibitem [{\citenamefont {Gadeyne}\ and\ \citenamefont {Ruchon}(2025)}]{gadeyne_2025_energy}%
  \BibitemOpen
  \bibfield  {author} {\bibinfo {author} {\bibfnamefont {T.}~\bibnamefont {Gadeyne}}\ and\ \bibinfo {author} {\bibfnamefont {T.}~\bibnamefont {Ruchon}},\ }\bibfield  {title} {\bibinfo {title} {Energy and photon centroids of spatiotemporal light pulses and consequences for their intrinsic orbital angular momentum},\ }\href {https://doi.org/10.1103/1wsz-48br} {\bibfield  {journal} {\bibinfo  {journal} {Physical Review A}\ }\textbf {\bibinfo {volume} {112}} (\bibinfo {year} {2025})}\BibitemShut {NoStop}%
\bibitem [{\citenamefont {Porras}(2023)}]{porras_2023_propagation}%
  \BibitemOpen
  \bibfield  {author} {\bibinfo {author} {\bibfnamefont {M.~A.}\ \bibnamefont {Porras}},\ }\bibfield  {title} {\bibinfo {title} {Propagation of higher-order spatiotemporal vortices},\ }\href {https://doi.org/10.1364/ol.479566} {\bibfield  {journal} {\bibinfo  {journal} {Optics Letters}\ }\textbf {\bibinfo {volume} {48}},\ \bibinfo {pages} {367} (\bibinfo {year} {2023})}\BibitemShut {NoStop}%
\bibitem [{\citenamefont {Porras}(2025)}]{porras_2025_optics}%
  \BibitemOpen
  \bibfield  {author} {\bibinfo {author} {\bibfnamefont {M.~A.}\ \bibnamefont {Porras}},\ }\bibfield  {title} {\bibinfo {title} {Optics of spatiotemporal optical vortices for atto- and nano-photonics},\ }\href {https://doi.org/10.1515/nanoph-2024-0544} {\bibfield  {journal} {\bibinfo  {journal} {Nanophotonics}\ } (\bibinfo {year} {2025})}\BibitemShut {NoStop}%
\bibitem [{\citenamefont {Gui}\ \emph {et~al.}(2022)\citenamefont {Gui}, \citenamefont {Brooks}, \citenamefont {Wang}, \citenamefont {Kapteyn}, \citenamefont {Murnane},\ and\ \citenamefont {Liao}}]{gui_2022_single}%
  \BibitemOpen
  \bibfield  {author} {\bibinfo {author} {\bibfnamefont {G.}~\bibnamefont {Gui}}, \bibinfo {author} {\bibfnamefont {N.~J.}\ \bibnamefont {Brooks}}, \bibinfo {author} {\bibfnamefont {B.}~\bibnamefont {Wang}}, \bibinfo {author} {\bibfnamefont {H.~C.}\ \bibnamefont {Kapteyn}}, \bibinfo {author} {\bibfnamefont {M.~M.}\ \bibnamefont {Murnane}},\ and\ \bibinfo {author} {\bibfnamefont {C.-T.}\ \bibnamefont {Liao}},\ }\bibfield  {title} {\bibinfo {title} {Single-frame characterization of ultrafast pulses with spatiotemporal orbital angular momentum},\ }\href {https://doi.org/10.1021/acsphotonics.2c00626} {\bibfield  {journal} {\bibinfo  {journal} {{ACS} Photonics}\ }\textbf {\bibinfo {volume} {9}},\ \bibinfo {pages} {2802} (\bibinfo {year} {2022})}\BibitemShut {NoStop}%
\bibitem [{\citenamefont {Gadeyne}(2026)}]{gadeyne_2026_spatiotemporal}%
  \BibitemOpen
  \bibfield  {author} {\bibinfo {author} {\bibfnamefont {T.}~\bibnamefont {Gadeyne}},\ }\emph {\bibinfo {title} {Spatiotemporal optical vortices and their transverse orbital angular momentum: propagation and high-order harmonics}},\ \href {https://doi.org/https://doi.org/10.70675/d891c471za48dz4f51zaf1ezc7918b6dbf4d} {Ph.D. thesis} (\bibinfo {year} {2026})\BibitemShut {NoStop}%
\bibitem [{\citenamefont {Trebino}(2000)}]{trebino_2000_frequency}%
  \BibitemOpen
  \bibfield  {author} {\bibinfo {author} {\bibfnamefont {R.}~\bibnamefont {Trebino}},\ }\href {https://doi.org/10.1007/978-1-4615-1181-6} {\emph {\bibinfo {title} {Frequency-Resolved Optical Gating: The Measurement of Ultrashort Laser Pulses}}}\ (\bibinfo  {publisher} {Springer US},\ \bibinfo {year} {2000})\BibitemShut {NoStop}%
\bibitem [{\citenamefont {Karimi}\ \emph {et~al.}(2007)\citenamefont {Karimi}, \citenamefont {Zito}, \citenamefont {Piccirillo}, \citenamefont {Marrucci},\ and\ \citenamefont {Santamato}}]{karimi_2007_hypergeometric}%
  \BibitemOpen
  \bibfield  {author} {\bibinfo {author} {\bibfnamefont {E.}~\bibnamefont {Karimi}}, \bibinfo {author} {\bibfnamefont {G.}~\bibnamefont {Zito}}, \bibinfo {author} {\bibfnamefont {B.}~\bibnamefont {Piccirillo}}, \bibinfo {author} {\bibfnamefont {L.}~\bibnamefont {Marrucci}},\ and\ \bibinfo {author} {\bibfnamefont {E.}~\bibnamefont {Santamato}},\ }\bibfield  {title} {\bibinfo {title} {Hypergeometric-gaussian modes},\ }\href {https://doi.org/10.1364/ol.32.003053} {\bibfield  {journal} {\bibinfo  {journal} {Optics Letters}\ }\textbf {\bibinfo {volume} {32}},\ \bibinfo {pages} {3053} (\bibinfo {year} {2007})}\BibitemShut {NoStop}%
\bibitem [{\citenamefont {Bekshaev}\ \emph {et~al.}(2009)\citenamefont {Bekshaev}, \citenamefont {Karamoch}, \citenamefont {Vasnetsov}, \citenamefont {Pas'ko},\ and\ \citenamefont {Soskin}}]{bekshaev_2009_structure}%
  \BibitemOpen
  \bibfield  {author} {\bibinfo {author} {\bibfnamefont {A.~Y.}\ \bibnamefont {Bekshaev}}, \bibinfo {author} {\bibfnamefont {A.~I.}\ \bibnamefont {Karamoch}}, \bibinfo {author} {\bibfnamefont {M.~V.}\ \bibnamefont {Vasnetsov}}, \bibinfo {author} {\bibfnamefont {V.~A.}\ \bibnamefont {Pas'ko}},\ and\ \bibinfo {author} {\bibfnamefont {M.~S.}\ \bibnamefont {Soskin}},\ }\href {https://doi.org/10.48550/ARXIV.0906.2619} {\bibinfo {title} {Structure of optical vortices produced by holographic gratings with "fork" geometry: Kummer beams}} (\bibinfo {year} {2009})\BibitemShut {NoStop}%
\bibitem [{\citenamefont {Vallone}(2015)}]{vallone_2015_properties}%
  \BibitemOpen
  \bibfield  {author} {\bibinfo {author} {\bibfnamefont {G.}~\bibnamefont {Vallone}},\ }\bibfield  {title} {\bibinfo {title} {On the properties of circular beams: normalization, laguerre–gauss expansion, and free-space divergence},\ }\href {https://doi.org/10.1364/ol.40.001717} {\bibfield  {journal} {\bibinfo  {journal} {Optics Letters}\ }\textbf {\bibinfo {volume} {40}},\ \bibinfo {pages} {1717} (\bibinfo {year} {2015})}\BibitemShut {NoStop}%
\bibitem [{\citenamefont {Self}(1983)}]{self_1983_focusing}%
  \BibitemOpen
  \bibfield  {author} {\bibinfo {author} {\bibfnamefont {S.~A.}\ \bibnamefont {Self}},\ }\bibfield  {title} {\bibinfo {title} {Focusing of spherical gaussian beams},\ }\href {https://doi.org/10.1364/ao.22.000658} {\bibfield  {journal} {\bibinfo  {journal} {Applied Optics}\ }\textbf {\bibinfo {volume} {22}},\ \bibinfo {pages} {658} (\bibinfo {year} {1983})}\BibitemShut {NoStop}%
\bibitem [{\citenamefont {Porras}\ and\ \citenamefont {Jolly}(2024)}]{porras_2024_procedure}%
  \BibitemOpen
  \bibfield  {author} {\bibinfo {author} {\bibfnamefont {M.~A.}\ \bibnamefont {Porras}}\ and\ \bibinfo {author} {\bibfnamefont {S.~W.}\ \bibnamefont {Jolly}},\ }\bibfield  {title} {\bibinfo {title} {Procedure for imparting transverse orbital angular momentum by focusing spatiotemporally coupled ultrashort pulses},\ }\href {https://doi.org/10.1103/physreva.109.033514} {\bibfield  {journal} {\bibinfo  {journal} {Physical Review A}\ }\textbf {\bibinfo {volume} {109}},\ \bibinfo {pages} {033514} (\bibinfo {year} {2024})}\BibitemShut {NoStop}%
\bibitem [{\citenamefont {Reintjes}(1984)}]{reintjes_1984_nonlinear}%
  \BibitemOpen
  \bibfield  {author} {\bibinfo {author} {\bibfnamefont {J.}~\bibnamefont {Reintjes}},\ }\href {https://doi.org/10.1016/b978-0-12-585980-6.x5001-2} {\emph {\bibinfo {title} {Nonlinear Optical Parametric Processes in Liquids and Gases}}}\ (\bibinfo  {publisher} {Academic Press},\ \bibinfo {year} {1984})\BibitemShut {NoStop}%
\bibitem [{\citenamefont {Boyd}(2008)}]{boyd_nonlinear_2008}%
  \BibitemOpen
  \bibfield  {author} {\bibinfo {author} {\bibfnamefont {R.~W.}\ \bibnamefont {Boyd}},\ }\href@noop {} {\emph {\bibinfo {title} {Nonlinear Optics}}},\ \bibinfo {edition} {3rd}\ ed.\ (\bibinfo  {publisher} {Academic Press, Inc.},\ \bibinfo {address} {USA},\ \bibinfo {year} {2008})\BibitemShut {NoStop}%
\bibitem [{\citenamefont {Martín-Hernández}\ \emph {et~al.}(2025{\natexlab{b}})\citenamefont {Martín-Hernández}, \citenamefont {Plaja}, \citenamefont {Hernández-García},\ and\ \citenamefont {Porras}}]{martinhernandez_2025_isolateda}%
  \BibitemOpen
  \bibfield  {author} {\bibinfo {author} {\bibfnamefont {R.}~\bibnamefont {Martín-Hernández}}, \bibinfo {author} {\bibfnamefont {L.}~\bibnamefont {Plaja}}, \bibinfo {author} {\bibfnamefont {C.}~\bibnamefont {Hernández-García}},\ and\ \bibinfo {author} {\bibfnamefont {M.~A.}\ \bibnamefont {Porras}},\ }\bibfield  {title} {\bibinfo {title} {Isolated attosecond spatiotemporal optical vortices: Interplay between the topological charge and orbital angular momentum scaling in high harmonic generation},\ }\href {https://doi.org/10.1103/kc1g-43mt} {\bibfield  {journal} {\bibinfo  {journal} {Physical Review Letters}\ }\textbf {\bibinfo {volume} {135}} (\bibinfo {year} {2025}{\natexlab{b}})}\BibitemShut {NoStop}%
\bibitem [{\citenamefont {Wan}\ \emph {et~al.}(2022)\citenamefont {Wan}, \citenamefont {Cao}, \citenamefont {Chen}, \citenamefont {Chong},\ and\ \citenamefont {Zhan}}]{wan_2022_toroidal}%
  \BibitemOpen
  \bibfield  {author} {\bibinfo {author} {\bibfnamefont {C.}~\bibnamefont {Wan}}, \bibinfo {author} {\bibfnamefont {Q.}~\bibnamefont {Cao}}, \bibinfo {author} {\bibfnamefont {J.}~\bibnamefont {Chen}}, \bibinfo {author} {\bibfnamefont {A.}~\bibnamefont {Chong}},\ and\ \bibinfo {author} {\bibfnamefont {Q.}~\bibnamefont {Zhan}},\ }\bibfield  {title} {\bibinfo {title} {Toroidal vortices of light},\ }\href {https://doi.org/10.1038/s41566-022-01013-y} {\bibfield  {journal} {\bibinfo  {journal} {Nature Photonics}\ }\textbf {\bibinfo {volume} {16}},\ \bibinfo {pages} {519} (\bibinfo {year} {2022})}\BibitemShut {NoStop}%
\bibitem [{\citenamefont {Zhou}\ \emph {et~al.}(2026)\citenamefont {Zhou}, \citenamefont {Zhang}, \citenamefont {Zhou}, \citenamefont {Zhang}, \citenamefont {Liu}, \citenamefont {Liang}, \citenamefont {Ponomarenko}, \citenamefont {Zhan}, \citenamefont {Cai},\ and\ \citenamefont {Liu}}]{zhou_2026_spatiotemporally}%
  \BibitemOpen
  \bibfield  {author} {\bibinfo {author} {\bibfnamefont {Y.}~\bibnamefont {Zhou}}, \bibinfo {author} {\bibfnamefont {N.}~\bibnamefont {Zhang}}, \bibinfo {author} {\bibfnamefont {A.}~\bibnamefont {Zhou}}, \bibinfo {author} {\bibfnamefont {Z.}~\bibnamefont {Zhang}}, \bibinfo {author} {\bibfnamefont {J.}~\bibnamefont {Liu}}, \bibinfo {author} {\bibfnamefont {C.}~\bibnamefont {Liang}}, \bibinfo {author} {\bibfnamefont {S.~A.}\ \bibnamefont {Ponomarenko}}, \bibinfo {author} {\bibfnamefont {Q.}~\bibnamefont {Zhan}}, \bibinfo {author} {\bibfnamefont {Y.}~\bibnamefont {Cai}},\ and\ \bibinfo {author} {\bibfnamefont {X.}~\bibnamefont {Liu}},\ }\bibfield  {title} {\bibinfo {title} {Spatiotemporally localized optical links and knots},\ }\href {https://doi.org/10.1038/s41467-026-72774-1} {\bibfield  {journal} {\bibinfo  {journal} {Nature Communications}\ }\textbf {\bibinfo {volume} {17}} (\bibinfo {year} {2026})}\BibitemShut {NoStop}%
\end{thebibliography}%

\end{document}